# PRODUCTION AND APPLICATION OF METAL-BASED NANOPARTICLES

*Vasilieva E.S., Tolochko O.V., Yudin V.E.[1], Kim D.[2], Lee D.-W.[2]*

Saint-Petersburg Polytechnical University, 195251, Russia, Saint-Petersburg, Polytechnicheskaya 29

[1]Institute of Macromolecules Research, 199004, Russia, Saint-Petersburg, V.O. Bolshoi Prospect 31.

[2] Nano Powder Metallurgy Group, Korea Institute of Machinery and Materials (KIMM) 66, Sangnam, Changwon, Kyungnam,641-010, Korea

## 1. ABSTRACT

A number of metal-based nanopowders such kinds as Fe, Co, Fe/Co alloy, Fe/C, Fe/organic shell were successfully produced by aerosol synthesis method. The mechanism of nanoparticles formation and the influence of experimental parameters on shape, size distribution, structure, chemical and phase composition of oxide-, carbon-, or organic- coated nanoparticles were evaluated. The sizes of particles can be varied from 6-100 nm with narrow size distribution. The several application fields of synthesized nanoparticles have been studied.

## 2. INTRODUCTION

In recent years, much attention has been paid to the synthesis and study of nanoparticles because of wide range of potential applications [1]. Particularly, magnetic nanoparticles can have the special characteristic of exhibiting single-domain magnetism and can be used in magnetic tapes, ferrofluid, magnetic refrigerants, etc, because of their ultra fine size dimensioned with magnetic domain size [2]. Oxide coated iron nanoparticles are non-toxic materials and due to enhanced magnetic properties can be more effective as compare with magnetite nanoparticles for biomedical application [3-4].

A wide range of techniques to fabricate nanoparticles has been developed rapidly over the past decades [5–7]. Among them, chemical synthesis of nanoparticles is a rapidly growing field because of its versatile applicability to almost all materials and high rate of production capability with little agglomeration. Since the properties of these nanoparticles are basically determined by their mean size, size distribution, external shape, internal structure, and chemical composition, the characteristics of powders must be controlled during the production of the nanoparticles [5].

Production of metal-based nanoparticles had been developed by gas phase synthesis through the pyrolysis of organometallic precursors. We adjust this method as the method for industrial synthesis of commercially available nanopowder of mean size of 6-100 nm. This method runs without vacuum and gives a relative high productivity of laboratory equipment.

This paper will devote to synthesis of the different kinds magnetic Fe-based nanoparticles with the shells of different chemical composition, such as the oxide, carbide or organic matters. It is a brief review of our works on magnetic particles production and application. The effect of processing parameters on microstructure, size distribution, chemical and phase composition, and morphology of nanoparticles were investigated.

## 3. EXPERIMENTAL

### 3.1. Nanoparticles production

We had produced nanoparticles by improved chemical vapor condensation method (CVC), which was adjusted for mass production of nanoparticles. The general scheme of CVC process is shown in Fig. 1. To produce nanoparticles, a carrier gas of high purity argon or helium is fed through a heated bubbling unit containing the metalloorganic precursor at the vaporization temperature. The flow of carrier gas entraining precursor vapor passed through the heated tubular reactor to the work chamber. The precursor decomposed in that reactor and condensed in the clusters or particles. In order to produce nanoparticles based on different metals or chemical compounds the suitable precursors should be used. For example, we used iron $Fe(CO)_5$, cobalt $Co_2(CO)_7$, or tungsten $W(CO)_6$ carbonyls for synthesis of Fe, Co and W-based nanoparticles, respectively.

The system includes independent vaporization units, connecting directly with the entrance of high-temperature reactor with two or more reaction zones. Zone temperatures were in the range of 230-1100°C, and were optimized for each experiment. Using accelerating gas as well as carrier gas was applied for changing precursor residential time in reaction zone and it concentration in gas phase. All particles were deposited on the surface in the work chamber, from which powders can be scrapped and collected.

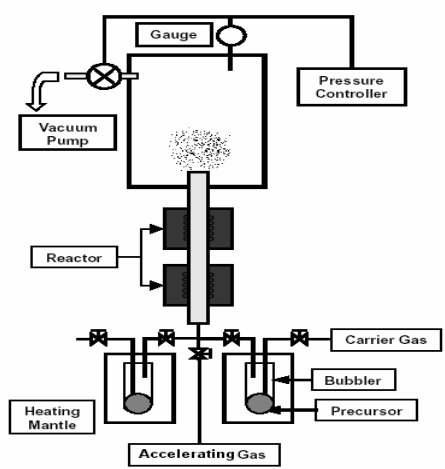

Fig.1. Scheme of equipment for method of aerosol synthesis of nanoparticles.

After exposure on air, usually metallic nanoparticles, specially iron, show spontaneous combustion. Hence, it is necessary to form nanoparticles with core-shell structure [8]. Shell cover metallic core and save it from oxygen atmosphere. Generally, oxide coated nanoparticles can be produced by passivation of as-produced nanoparticles in the mixing gas of Ar contained a trace $O_2$ flowed for 10h with pressure of $1kg/cm^2$ . That has slowly oxidized nanoparticles surface and than particles can be used on atmosphere air.

However, other kind of shells also can be created, such as carbide, noble metals, polymer, etc., by using the same experimental setup. The presence of two reaction zones allowed providing successive decomposition of precursors vapor in the same gas flow. If different chemical compound are used as the precursors for core and shell, usually temperature regime of it decomposition is also different. Therefore, in the gas phase, the formation of solid clusters of one component starts at first zone of reactor, than it follows to the next zone of higher temperature for decomposition another precursor. By that way as-formed clusters or nanoparticles can be covered by different substances.

### 3.2. Synthesis of nanoparticles-polymer composites

The polyimide based thin film magnetic composites were synthesized and investigated. At the first step the nanoparticles were ultrasonically dispersed in solvent, where all necessary for synthesis components were also added. Then the formed solution of forpolimer (polyamidacid (PAA)) with dispersed nanoparticles had been spilt on flat glass substrates. After drying the dense elastic films with the thickness of about 40 μm were obtained. Finally those films were heat-treated in vacuum by three steps regime: $100^oC/1h$, $200^oC/1h$ and $250^oC/1h$.

### 3.3. Characterization methods

The synthesized powders were characterized for crystal structure, phase composition, volume fraction of phases, particle size, and chemical impurities by using X-ray diffraction (XRD) method, scanning and transmission electron microscopy, thermogravimetric analysis, Mössbauer spectroscopy, X-ray photoelectron spectroscopy. The carbon content of the iron nanoparticles and Fe/Co ratio in the alloyed nanoparticles were determined by chemical elemental analysis. Magnetization of nanoparticles and polyimide composite samples was measured by a vibrating sample magnetometer (VSM) at room temperature in a field up to 10 kOe.

## 3. EXPERIMENTAL RESULTS AND DISCUSSION

### 3.1. Oxide-coated iron nanoparticles

Synthesis and characterization of oxide-coated iron nanoparticles was described previously [7-9]. With the increase of reaction temperatures, the average size of the iron particles increased and the particle size distribution becomes wider and more asymmetrical with the size increment. Particles did not appear at the temperatures lower than 230°C. The shells consisted of magnetite [8] and their thickness was evaluated about 2-3nm. X-ray diffraction pattern didn't show any additional phase except BCC iron; however, slight diffusive peaks which belonged to oxides were distinguished. Mossbauer spectroscopy shows that according to CVC parameters, particles' structure contains about 15-100% of bulk iron and up to 45% of $Fe_3O_4$, another phases can be Fe (Fine), $\alpha$-FeOOH (Fine and middle), or $\gamma$-FeOOH.

By variation of experimental parameters such as precursor decomposition temperature, carbonyl evaporation rate and carrier gas flow rate, the mean particles size can be changed from 6 to 150 nm [10]. At the low temperature, particles have round shape and core-shell structure, core is metallic iron and shell can be crystalline or amorphous oxides. The increasing decomposition temperature and residential time of nanoparticles in reactor increased the mean particles size and lead to more asymmetric size distribution. Unagglomerated magnetic nanoparticles form intricate long threads to minimize the magnetic energy. At the higher temperature, particles coagulated and structure completely consisted of filaments of 40-100nm in diameter depending on experimental parameters. Such structure is so unstable that new large particles are formed in the temperature of higher than 700°C and does not grow significantly as the temperature increases up to 1000°C.

Fig. 2 shows the magnetic properties of the iron nanoparticles synthesized in various conditions. Maximal magnetization of the powder with the average particle size of 75 nm is about 210 emu/g, which is almost theoretical value of 225 emu/g known for pure bulk iron. At the smaller mean size of particles, magnetization is also in the good correspondence with values which were calculated in speculation that particles consist of iron core and magnetite shell of 2 or 3nm thickness ($M_s$ of magnetite ~90emu/g). However, when the particle size is smaller than 12nm the maximum magnetization of the powder decreases continuously. Such decrease of magnetization comes from the increase of portion of superparamagnetic particles with the decrease of mean particle size. Particles smaller them 8 nm behave like paramagnets.

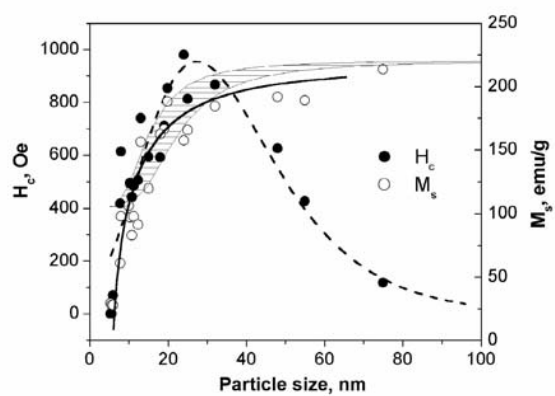

Fig. 2. Coercive force ($H_c$) and maximal magnetization ($M_s$) with particle size

That is common, that the coercivity of ferromagnetic materials increases with the decrease of particle size until it is maximized to the size of single magnetic domain. The maximum coercivity obtained so far is about 1 kOe, measured from the iron nanopowder with the average particle size of 20-25 nm.

### 3.2. Iron-cobalt nanoparticles with improved magnetic properties

To produce alloyed Fe-Co metallic nanoparticles the evaporation of precursors was organized in independent vaporization units and precursors vapor concentration in carrier gas was controlled. Joint decomposition of iron and cobalt precursors was done at the same pyrolytic temperature in reactor. Oxide shells were formed by the said passivation process.

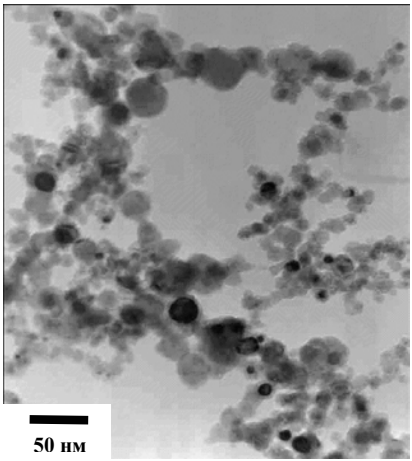

Fig.3. TEM microphotographs of iron-cobalt alloyed nanoparticles.

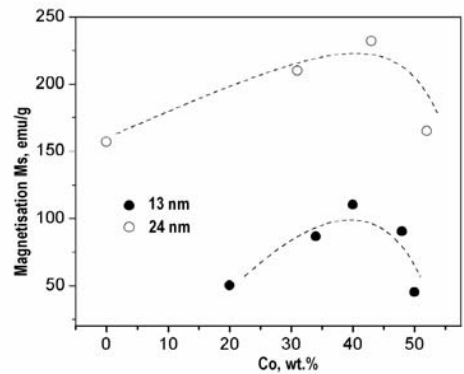

Fig.4. Saturation magnetization of iron-cobalt alloyed nanoparticles versus cobalt content.

Particles have approximately spherical shape with mean size in range of 13-60 nm. TEM images (Fig.3) shows non agglomerated particles with core-shell structure and the typical morphology of passivated Fe-based nanoparticles [8,10]. XRD analysis shows single phase BCC solid solution.

Fig. 4 shows the Co content dependence of the saturation magnetization of the Fe-Co particles. Although a variation in $M_s$ can be induced by a change in the average particle size. With increasing Co content the saturation magnetization increases and reaches its maximum at near 40 wt% Co, and then decreases and coercive force continuously increases and have maximal measured value of 1250 Oe.

### 3.3. Carbon-coated Fe-based nanoparticles

To produce Fe and Fe-C nanoparticles, a high purity carbon monoxide (CO) was used as a source of carbon and as a carrier gas. It is known that carbon appears through the reaction of CO disproportionation reaction: $2CO_g = C_s + CO_{2\,g}$, $\Delta H$=169 kJ/mol. To run that reaction metallic catalytic surface is necessary. In order to study the preferable conditions for that reaction, the kinetic investigations using 15 nm sized iron particles were carried out. The appreciable reaction rates are in the temperature interval from 470 to 820 °C with a maximum rate at 625°C [11]

This reaction occurs also even at lower temperatures (down to 325°C) and temperatures up to 1000°C however at very low rates.

In order to investigate the effect of experimental parameters on particle size, phase composition and structure, several particle samples were synthesized at various pyrolytic temperatures from $400^0$C to $1100^0$C. The increasing of gas flow rate decreases particle size and leads to the changes in particle phase composition as shown in the Fig.5.

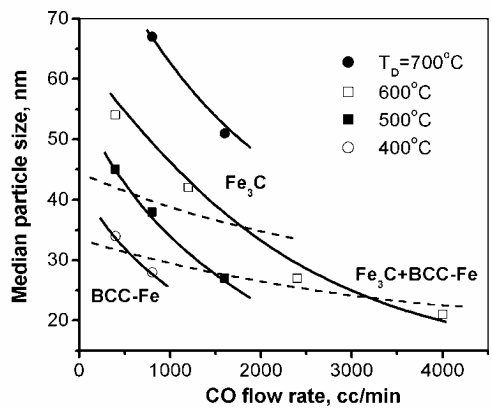

Fig.5. Phase composition and particle size versus gas flow rate at different decomposition temperatures ($T_D$).

Particles have core-shell structure. XPS and TEM study of shells shows that the shell structure depends on reactor temperature. At the lowest temperatures (400 and 500°C) the shell was soot and at the higher temperatures it was graphite. At the temperatures higher than 1100°C the fullerene –like structure of shells and single walled carbon nanotubes can be formed [12].

### 3.2. Organic coated magnetic nanoparticles

In case of synthesis of surface-modified iron nanoparticles, the organic medium with chemical composition of $R_{fo}$-S-S-S-$R_{fo}$, where S is sulphur and $R_{fo}$ is radical of $CF_3OCF_2CF_2CF_2OCF_2CF_2$, had been specially synthesized and applied for nanoparticles covering. It has evaporation temperature of 100°C and decomposition temperature higher that 250°C. The decomposition run through the sulphur links. After decomposition it has opened sulphur link strongly connected through the sulphur with perfluorineoxide radical. Organic-coated nanoparticles were produced by decomposition of iron carbonyl with iron nanoparticles formation at 230°C and it subsequent reaction with $R_{fo}$-S- vapor, which had connected with iron through the Fe-S link and covered as-formed iron nanoparticles by $R_{f0}$ radicals.

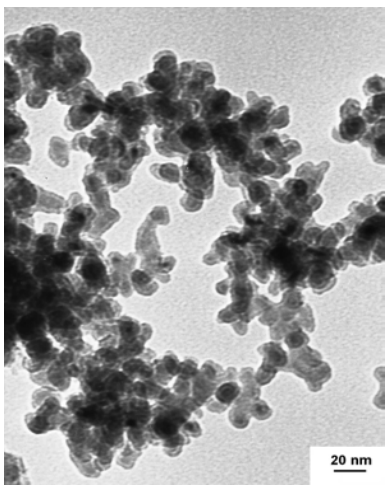

Fig.6. TEM microphotographs of organic coated iron nanoparticles.

Thus, it were synthesized low agglomerated nanoparticles with core-shell structure of several mean size ranged of 8-15 nm as it was found from TEM results (Fig. 6). Nanoparticles have quite narrow size distribution.
Analysis of phase composition by XRD and Mossbauer spectroscopy methods shows presence of pure α-Fe and sulphude phases and the absence of the traces of oxides. However, the values of Mossbauer spectra parameters, such as quadruple splitting and isomer shift, are higher as compare with Fe-S compounds. It can mean that organic radicals weight down of electronically density per sulphur from iron nuclei.

Produced particles shows the better damping in organic mediums, specially in compatible with shell medium, hereby stable of suspension in time was significantly improved. Such magnetic fluids can be applicable for machinery as well as for medical diagnostic and treatment.

### 3.3. Polymer-based composites

Flexible polyimide-based films were synthesized at the nanoparticles content was changed from 1 to 20 wt.%. With increasing particles weight up to 30 wt.% we couldn't get uniform distribution of particles in matrix and flexible samples wasn't achieved. Samples were optically transparent at the particles content up to 10 wt.%. With increasing particles content the composite film have about the same tensile strength of 105±2 MPa. It is important to notice that Young modulus increase from 2,4 to 3,0 GPa with increasing particles content from 0 to 20wt.%, respectively. Micrograph of ached nanoparticles filled polyimide is shown in the Fig.7.

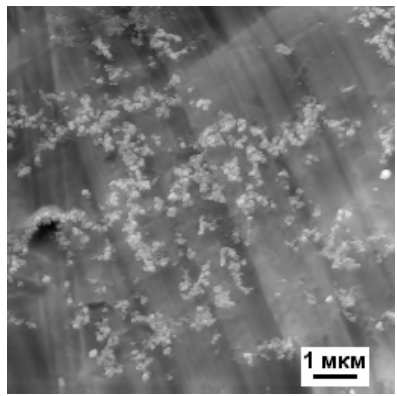

Fig. 7. Structure of nanoparticles-polymer composite.

The measurements of composite magnetic properties show linear growth of saturation magnetization, which was proportional to particle content in polymer. However, coercive force doesn't show any dependence from the content of nanoparticles and in every sample was significantly smaller as compare with coercivity of initial particles.

In order to improve magnetic properties of composite sample the constant transversal magnetic field with value 0,01 T had been applied to forpolymer (PAA) film sample during of drying process. Hence magnetic moments of single-domain nanoparticles have to be oriented in thin films accordingly of field direction and magnetic structure will be ordered. Produced composite samples shows coercivity of 970 Oe, which is close to coercivity of initial Fe-Co alloyed particles (1037 Oe). Saturation magnetization shows value of 58 emu/g at the Fe-Co alloyed particles content of 20 wt.%. That is in a good correspondence with value of initial filler nanopowder (240 emu/g).

## 4. CONCLUSION

The possibility to produce multicomponent, alloyed nanoparticles and nanoparticles of chemical compounds by gas phase synthesis method was shown. Iron-based magnetic nanoparticles, such as oxide-coated iron, $Fe_3C$, Fe-Co alloyed, organic and carbon-coated iron nanoparticles of 6-75nm were successfully produced. Size distribution, phase and chemical composition of nanoparticles and shells can be precisely controlled by experimental parameters.

It is possible to produce superparamagnetic nanoparticles and nanoparticles with $H_c$ of ~ 1.2 kOe, and high magnetization (up to 245 emu/g), which would be applicable "in-vivo" for magnetic resonance imaging as contrast-increasing materials, for drug delivery, and for experiments on magnetic hyperthermia, as same as for magnetic recording media, permanent magnets and magnetic plastics.

## 5. REFERENCES

[1] L. Hu, M. Chen. Preparation of Ultrafine Powder: the Frontier of Chemical Engineering (review) // Materials Chemistry and Physics. 1996. V.43. P.212– 219.

[2] G.C. Hadjipanayis, G.A. Prinz, Science and Technology of Nanostructured Magnetic Materials, Plenum, New York, 1991.

[3] Y. Zhang, N. Kohler, M. Zhang Surface modification of superparamagnetic magnetite nanoparticles and their intracellular uptake. Biomaterials 2002;23:1553–61.

[4] O. Bomatr -Miguel, Marra P. Morales, Pedro Tartaja, Jesurs Ruiz-Cabellob, Pierre Bonvillec, Martərn Santosd, Xinqing Zhaoe, Veintemillas-Verdaguera S. Fe-based nanoparticulate metallic alloys as contrast agents for magnetic resonance imaging. Biomaterials 2005; 26: P. 5695–5703.

[5] T. Sugimoto (Ed.), Fine Particles—Synthesis, Characterization and Mechanism of Growth, Marcel Dekker, New York, 1996.

[6] Ultra-Fine Particles: Exploratory, Science and Technology ed. by C. Hayashi, R.Ueda, A.Tasaki. 1997. Noyes Publ. Westwood, NJ. USA. 447 p.

[7] S. Veintemillas-Verdaguer, M.P. Morales, C.J. Serna, Continuous production of g-Fe2O3 ultrafne powders by laser pyrolysis. Mater. Lett. 35 (1998) 227– 231.

[8] C.J. Choi, O. Tolochko, B.K. Kim Preparation of iron nanoparticles by chemical vapor condensation. // Materials Letters 2002, 56: P 289– 294

[9] D.W. Lee, T.S. Jang, D. Kim, O.V. Tolochko, B.K. Kim Nano-Crystalline Iron Particles Synthesized Without Chilling by Chemical Vapor Condensation // Glass Physics and Chemistry 2005 V.31. No4. Pp.545-548.

[10] D. Kim, E.S. Vasilieva, A.G. Nasibulin, D.W. Lee, O.V. Tolochko, B.K. Kim. Aerosol Synthesis and Growth Mechanism of Magnetic Iron Nanoparticles // Mater. Sci. Forum Vols 534-536 (2007). 9. (on-line at www.scientific.net)

[11] E.S.Vasilieva, A.G. Nasibulin, D.W. Lee, O.V. Tolochko, E. I. Kauppinen. Synthesis of nanoparticles by the gas phase decomposition of iron pentacarbonyl in the carbon monoxide atmosphere // Phys.-Chem. Kinetics in the Gas Dynamics. 2006. V. 4. 9 on-line at: www.chemphys.edu.ru/pdf/2006-07-18-001.pdf. (in Russian)

[12] A.G. Nasibulin, P. Queipo, S. D. Shandakov, D.P. Brown, H. Jiang, P.V. Pikhitsa, O.V. Tolochko, and E.I. Kauppinen. Studies on Mechanism of Single-Walled Carbon Nanotube Formation // Journal of Nanoscience and Nanotechnology. 2006. V.6. 1-14.